\documentclass{article}

\usepackage[english]{babel}

\usepackage[letterpaper,top=2cm,bottom=2cm,left=3cm,right=3cm,marginparwidth=1.75cm]{geometry}
\usepackage{authblk}
\usepackage[utf8]{inputenc}
\usepackage{amsmath}
\usepackage{dsfont}
\usepackage{amsfonts}
\usepackage{mathrsfs}
\usepackage{amsmath}
\usepackage{graphicx}
\usepackage[colorlinks=true, allcolors=blue]{hyperref}

\title{New results concerning the curvature of the Universe}

\begin{document}

\author[1]{Maxim Khlopov\thanks{Email: \texttt{khlopov@apc.in2p3.fr}}}
\author[2,3]{Maxim Krasnov\thanks{Email: \texttt{morrowindman1@mail.ru}}}
\author[4]{Jan Novák\thanks{Email: \texttt{jan.novak@johnynewman.com}}}

\affil[1]{Virtual Institute of Astroparticle Physics, 75018 Paris, 
        France}
\affil[2]{National Research Nuclear University MEPhI, Moscow, 115409, Russia}
\affil[3]{Research Institute of Physics, Southern Federal University, Rostov-on-Don, 344090, Russia}
\affil[4]{Department of Physics, Faculty of Mechanical Engineering, Czech Technical University in Prague, Prague, 166 07, Czech Republic}

\date{\today}

\maketitle 

\begin{abstract} 
One of the fundamental questions we ask in cosmology is what the spatial curvature of the universe is. Experimental observations show that the universe is very close to being spatially flat. However, new results, mainly from measurements by the Planck satellite, indicate that the curvature parameter might be positive. We will, however, examine completely new theoretical results that suggest that we actually live in a closed Universe.
\end{abstract} 

\section*{Introduction}
The curvature of the Universe describes its overall geometry on a cosmic scale, which is determined by the total density of matter and energy within it. Discussions about what the actual value of the curvature parameter $k$ is have been going on since the beginning of modern cosmology in the 1920s. The reason, of course, is that this parameter appears in the fundamental Friedmann equations, and it was also immediately known that there exist three basic types of geometries of the space around us.

The first possibility---very straightforward---is that we live in a flat universe (this geometry is also called by other terms, Euclidean, when the curvature parameter is zero). In this geometry, it holds that the sum of the angles of a big triangle in the Universe is $180^0$ and that a circumference of a circle of radius $r$ is exactly $2\pi r$. A flat universe is infinite in extent and actually forms a kind of limiting case for other two kinds of "shape" of the space. 

This is primarily a closed universe (spherical with positive curvature), where parallel lines eventually converge. The sum of the angles of a triangle is more than $180^0$ and the circumference of a circle of radius $r$ is less than $2\pi r$. We can vividly imagine it using a sphere, which has a finite surface but no boundary. If we were to set out from a certain point in a given direction, we would eventually return to that point from the other side. 

And finally, an open universe (hyperbolic with negative curvature), where we could draw more lines parallel to a given line through a given point, or in other words, parallel lines diverge. A triangle's angles sum to less than $180^0$ and the circumference of a circle of radius $r$ is more than $2\pi r$. It is infinite in volume, and this geometry can be imagined according to a riding saddle. 

The primary importance of knowing the curvature parameter $k$ arises from the fact that, according to the Friedmann equations (without the cosmological constant), if the universe has just the critical density---that is, if it is spatially flat---it will continue to expand forever (but increasingly slowly). If the density is greater than the critical density, the expansion will eventually stop, and the universe will collapse into a Big Crunch. If the density is less than the critical one, it will continue to spread forever. Today we already know that the situation is not so simple as described according to the initial Friedmann models, because at this moment the universe is expanding with acceleration, for which some dark energy is responsible, which in the simplest case is modeled using the cosmological constant in the Friedmann equations. However, the question of critical density still remains crucial, \cite{Liddle}. 

From the experimental point of view, it was believed for a very long time that the universe is spatially flat. But now --- according to the results of the Planck satellite, everything indicates that the Universe is closed, \cite{bouchert, Ivanov, Shen, Campo, Handley} (model-dependent finding). And in this article, we wish to point out that this is completely in accordance with modern theory. Primarily from work \cite{Easson, EassonD} follow no-go theorems: spatially flat ($k = 0$) and open ($k = -1$) Friedmann-Lemaitre-Robertson-Walker  spacetimes cannot be simultaneously non-singular, geodesically complete, and consistent with the
averaged null energy condition (ANEC); any non-static flat or open universe that is complete must therefore violate the ANEC. Only $k = +1$ universes actually admit such
models, with global de Sitter space providing the canonical realization that saturates the bound. So, positive spatial curvature emerges as a fundamental geometric ingredient of non-singular cosmology.

Further, we recently formulated an approach to quantum gravity --- the ring paradigm (RP), \cite{JNa1, JNa2, JNa4, JNa5}. According to this theory, gravity is an extremely non-local interaction. The graviton is modeled as a phonon on a lattice of matter in the universe. And what is very important, after applying it to cosmology, we obtain a solution to the old problem of the cosmological constant. (Preferred models are also cyclic models of the universe, as we show in one of our recent works, \cite{JNa6}.) However, RP can only be formulated in a closed universe (because of the ultimate speed of creating gravitational rings), which is another strong argument that our universe has a geometry similar to the geometry on a sphere. And it is not without interest that even the no-boundary proposal of Hawking and Hartle - which is again based on computation in quantum cosmology - cannot be formulated in a flat or open universe, \cite{Khlopov}. 

This work will be organized as follows: first we present the experimental evidence for a nearly flat universe---or for the closed universe---and then we introduce new theoretical arguments in favor of the closed universe.

\section*{Experimental observation of the curvature of the universe}





The spatial curvature of the Universe, quantified by the curvature density parameter $\Omega_{k0}$, is a fundamental observable in cosmology. Its measurement provides a critical test of the inflationary paradigm, which generically predicts a flat universe, and serves as a cornerstone for the standard $\Lambda$CDM model. Over the past two decades, a variety of experimental probes have been employed to constrain $\Omega_{k0}$ (actually, $\Omega_k<0$ corresponds to a spatially closed universe), with results that have evolved from order-unity uncertainties to sub-percent precision. Here we will present a short review of modern approaches to constrain the curvature of the Universe.

\subsection*{Cosmic Microwave Background}

The Cosmic Microwave Background (CMB) has historically provided the strongest and most direct constraints on cosmic curvature. The angular scale of the acoustic peaks in the CMB power spectrum is sensitive to the geometry of the Universe through the angular diameter distance to the last scattering surface.

Early groundbreaking measurements from the BOOMERanG experiment provided some of the first strong evidence for a flat universe, with their results being consistent with $\Omega_{k0} \sim 0$~\cite{boomerang}. Further, the current CMB constraints come from the Planck satellite. The 2018 Planck results, when combined with temperature, polarisation, and lensing data, yield a measurement of $\Omega_{k0} = -0.0106 \pm 0.0065$~\cite{planck2018}, which is consistent with a flat universe. The Planck lensing amplitude has been found to be larger than expected within the standard $\Lambda$CDM model~\cite{planck2018, calabrese2008}, which can introduce a degeneracy with curvature. In fact, when the lensing data are omitted, the Planck constraints shift to favour a closed universe, with $-0.095 < \Omega_{k0} < -0.007$ at 99\% CL~\cite{divalentino2019, handley2021}. Therefore, CMB measurements are model-dependent, relying on assumptions about both the early and modern Universe.


To overcome the model dependence of the CMB, numerous studies have focused on constraining the curvature using modern Universe observables. These methods often aim to be more model-independent, relying on direct measurements of the expansion rate and distance scales.

\subsection*{Cosmic Chronometers and Supernovae}

A popular approach involves combining Cosmic Chronometers (CC) data, which provide a direct measurement of the Hubble parameter $H(z)$, with Type Ia Supernovae (SN) data, which measure luminosity distances. By comparing these two independent distance measures, one can constrain $\Omega_{k0}$ without assuming a specific dark energy models or early-universe physics.

Several works have pioneered this method. For instance, Sapone, Majerotto, and Nesseris~\cite{sapone2014} performed an early analysis testing the FLRW cosmology with this approach. Subsequent studies by Cai, Guo, and Yang~\cite{cai2016}, Li et al.~\cite{li2016}, and Jesus et al.~\cite{jesus2020} have refined these constraints. More recent analyses, such as those by Cao, Ryan, and Ratra~\cite{cao2021} using the Pantheon and DES supernova samples, and by Dhawan, Alsing, and Vagnozzi~\cite{dhawan2021} using non-parametric methods, have found results consistent with a flat universe. A comprehensive model-independent study by Favale, Gómez-Valent, and Migliaccio~\cite{favale2023} further solidified these findings, showing that the combination of CC and SN data strongly supports a spatially flat geometry.

\subsection*{Baryon Acoustic Oscillations}

Baryon Acoustic Oscillations (BAO) provide a standard ruler that can be used to measure the angular diameter distance and the Hubble parameter as a function of redshift. When combined with other data, BAO measurements offer a way to constrain curvature.

Yu, Ratra, and Wang~\cite{yu2018} and Ryan, Doshi, and Ratra~\cite{ryan2018} used BAO data in conjunction with Hubble parameter measurements to constrain $\Omega_{k0}$ and dark energy dynamics, finding consistency with a flat universe. More recently, the combination of BAO data from surveys like BOSS, eBOSS, and the first data release from DESI has been used to provide some of the most stringent late-universe constraints on curvature. For instance, Liu et al.~\cite{liu2025} introduced a method to measure $\Omega_{k0}$ independent of the sound horizon and $H_0$, achieving bounds of $|\Omega_{k0}| < 0.01$.



\subsection*{Gravitational Waves as Standard Sirens}

Gravitational-wave (GW) standard sirens offer an independent way for measuring cosmic distances, and consequently, the curvature of the Universe. Unlike standard candles such as Type Ia supernovae, which require an electromagnetic calibration of their luminosity, GWs from compact binary coalescences provide a direct, self-calibrated measurement of the luminosity distance from the waveform alone. When an electromagnetic counterpart is detected, the redshift \(z\) of the source can be determined, creating a standard siren.

This method is advantageous for model-independent curvature tests. By comparing the distance measurements from GW sources with independent measurements of the Hubble parameter from cosmic chronometers, one can constrain the spatial curvature \(\Omega_{k0}\) without assumptions about specific dark energy models or early-universe dynamics.

Wei~\cite{wei2018} pioneered this approach, simulating GW data from future detectors like the Einstein Telescope (ET). Their results showed that with 100 simulated GW events and 31 current cosmic-chronometer measurements, the curvature parameter \(\Omega_{k0}\) could be constrained at the level of \(\sim 0.125\). With 1000 GW events, the uncertainty would be reduced to \(\sim 0.040\). By additionally adding 50 mock \(H(z)\) data, the constraint could reach \(\Omega_{k0} = -0.002 \pm 0.028\), which is significantly tighter than many other model-independent methods using electromagnetic data.

Subsequent studies have explored the potential of future space-based detectors. Zheng et al.~\cite{zheng2021} investigated the capabilities of the DECi-hertz Interferometer Gravitational-wave Observatory (DECIGO). They found that DECIGO, acting as a standard siren, could constrain cosmic curvature with a precision of \(\Delta \Omega_{k0} = 0.09\) when combined with cosmic chronometer data. More interestingly, a model-independent Gaussian process reconstruction allows to perform the measurement of the evolution of the curvature parameter \(\Omega_k(z)\) as a function of redshift, which is beyond traditional electromagnetic methods.

Zhang et al.~\cite{zhang2022} further extended this analysis by incorporating the phase shift in the gravitational waveform caused by the accelerating expansion of the Universe. This additional observable allows for a direct measurement of the acceleration parameter, providing an independent way to determine the curvature in the GW domain. Their results indicate that DECIGO could provide a constraint on the cosmic curvature at a precision comparable to existing electromagnetic data, without strong evidence for a deviation from a flat universe.


\subsection*{The Alcock--Paczyński Effect}

The Alcock--Paczyński (AP) effect is a geometric distortion in the observed clustering of galaxies that arises from the wrong calculation of the Hubble parameter. 
However, if the assumed cosmology is wrong, the measured radial and transverse clustering scales will be mismatched, which will lead to an apparent anisotropy in the power spectrum. This distortion is quantified by the AP parameter, which depends on the ratio of the angular diameter distance \(D_A(z)\) to the Hubble parameter \(H(z)\).

Amendola, Marinucci, and Quartin~\cite{amendola2025} have recently proposed a novel methodology to measure the cosmological spatial curvature \(\Omega_{k0}\) by exploiting the deviation from statistical isotropy caused by the AP effect in large-scale galaxy clustering. It is independent of the calibration of standard candles, rulers, or clocks; independent of the power spectrum shape (and therefore of pre-recombination physics); independent of galaxy bias; independent of the theory of gravity; independent of the dark energy model; and independent of the background cosmology in general.

This method works by combining measurements of the galaxy power spectrum and bispectrum (the Fourier transform of the three-point correlation function). The power spectrum provides the standard AP distortion, while the bispectrum breaks degeneracies that would prevent a clean measurement of curvature. The authors find that a combined Dark Energy Spectroscopic Instrument (DESI) and Euclid galaxy survey could achieve a precision of
\begin{equation*}
\Delta \Omega_{k0} = 0.057 \quad (1\sigma \text{ C.L.})    
\end{equation*}
in the redshift range \(z < 2\).

This level of precision is competitive with current CMB constraints and, crucially, is achieved through a completely independent and model-agnostic method. The AP effect does not rely on the distance ladder or the sound horizon scale, making it an excellent cross-check for potential systematics in other probes. As the authors note, this technique provides a way to measure curvature that is independent of calibration of standard candles, rulers, or clocks, of the power spectrum shape (and thus also of the pre-recombination physics), of the galaxy bias, of the theory of gravity, of the dark energy model and of the background cosmology in general~\cite{amendola2025}.\\


So, the conclusion from this experimental part of our work is that most of the experiments point out the fact that the spatial curvature of the universe is very close to a flat universe, with a slight preference for models with positive curvature. However, we will show some fascinating theoretical arguments for why the geometry of the Universe should be the geometry of a sphere. 

\section*{Theoretical evidence for positive curvature}

It was proven by Penrose in 1965 that when a dying star collapses, \cite{Penrose}, it eventually forms a trapped surface. This is an invisible point of no return in space. Once matter or light crosses it, it cannot escape. In Einstein's general relativity, the presence of the trapped surface leads to the creation of a singularity --- a point where gravity becomes infinite and physics breaks down \footnote{Actually, we need to precisely distinguish that a trapped surface is a local or quasi-local spatial area where outgoing light rays are forced to converge, whereas an event horizon is a global, teleological boundary dividing spacetime into what could and could not escape to infinite distance.}. This is a point where the gravitational pull is so strong that even outgoing light rays are bent. Penrose's work mathematically proved that once this point is passed, nothing can stop the matter from collapsing completely into the central singularity, \cite{Gundlach, Bambi}. Actually, the singularity theorems demonstrate that, under broad and physically reasonable assumptions---including the validity of classical energy conditions and the presence of trapped surfaces---generic spacetimes in general relativity are geodesically incomplete, \cite{Ellis}. We interpret these results in a way that cosmological spacetimes contain a past singularity, such as a Big Bang singularity, \cite{Elizalde}, although the theorems technically establish only the existence of incomplete geodesics and not necessarily a singularity in
curvature or energy density.

A related but a little bit distinct result is the Borde-Guth-Vilenkin (BGV) theorem, which deals with the past completeness of inflationary spacetimes, \cite{Vilenkin}. We claim in the BGV theorem that any universe which has - on average -
been expanding along a past-directed geodesic must be
geodesically incomplete to the past, regardless of the details of the energy content or the validity of classical energy conditions. But unlike the Hawking-Penrose theorems,
the BGV result does not require the presence of trapped
surfaces or any specific matter model; it just relies only on the
assumption of positive average Hubble expansion along
timelike or null geodesics. The consequence is that the theorem is
often interpreted as suggesting that inflationary spacetimes, even if eternal into the future, could not be extended
indefinitely into the past within a classical spacetime description. The theorem, however, admits certain kinds of 
loopholes, and examples of past-eternal inflationary models are common \cite{Maartens, Murugan, Lesnefsky, EL, LE}.

When we want to formulate the completely new result from the paper \cite{Easson}, we need to explain what are ANEC (average null energy condition). It occupies a special place at the intersection of cosmology, disciplines studying the gravitation, and quantum field theory, because it asserts that the integral of
the null-null component of the stress-energy tensor along
any complete null geodesic is non-negative,
\begin{gather}
\int_{-\infty}^{\infty} T_{\mu\nu}k^{\mu}k^{\nu}\ d\lambda \geq 0.
\end{gather}
Unlike the classical pointwise energy conditions, which
could be routinely violated by quantum fluctuations, the
ANEC includes the weaker statement that negative energy densities cannot persist with
out compensation. It expresses the idea that---although
quantum fields may locally exhibit negative energy---the
total energy flux measured along a light ray must remain
positive in any consistent, unitary theory.

To be more explicit, we define the metric 
\begin{gather}
ds^2 = -dt^2 + a(t)^2d\Sigma_k^2,\ \ k\in\{0,\pm 1\}\nonumber
\end{gather}
with $a\in\mathscr{C}^2$ and $a(t)>0$. For a perfect fluid source, we define the affine ANEC integral via two-sided improper truncations 
\begin{gather}
 I_{ANEC} \equiv \lim_{T_{\pm}\rightarrow\pm\infty}\int_{T_{-}}^{T_{+}}\frac{\rho(t)+p(t)}{a(t)}\ dt.  
\end{gather}
We refrain from separating boundary and bulk limits unless each converges individually. Then we stated the theorem: 
{\it let $(M,g)$ be a non-static FLRW spacetime with $k=0$ or $k=-1$, $a(t)\in\mathscr{C}^2$, $a(t)>0$ and bounded curvature invariants. If $(M,g)$ is null geodesically complete, then $I_{ANEC}<0$. Hence, any $ANEC$-satisfying ($I_{ANEC}\geq 0$) flat or open FLRW ST is null-incomplete.} 

This theorem has far-reaching consequences because it prefers the closed universe, \cite{Damien}. One could say that we don't live in a FLRW universe, but we are---according to the data---with $\Lambda$CDM-model very close to it, so this result is relevant.\\  

But the most important argument pointing to the fact that we live in a closed universe comes from quantum gravity. We formulated a novel approach to quantization of gravity in \cite{JNa1, JNa2, JNa4, JNa5, JNa6}, RP. It is a Lorentz-breaking theory, where all the particles of the Standard Model are modeled as in string theory. Only the graviton on the full non-perturbative level is described differently as a fundamental 1-dimensional ring. This object is longitudinally vibrating, and what we call today a graviton in the perturbative sense is just a phonon, so quasi-particle. 

This theory should have a whole number of consequences for many areas of theoretical physics. Let's mention first again the singularity theorems, \cite{Penrose}. We claim in them that under very general assumptions the singularities in black holes form (spacetime must become geodesically incomplete). However, RP comes with the fact that there exists a phantom field in Nature, \cite{JNa1, JNa2, JNa4, JNa5, JNa6}. The existence of this field frequently leads just to the violation of the null-energy conditions in the singularity theorems. And further, the grid of rings---we refer to the basic postulates of this theory---is ever-present in the core of black holes and at the Big Bang and it shrinks always to some minimal volume. This means that RP is a candidate for a non-singular theory. 

This is related to the fact that there must exist some pre-Big Bang phase in the development of the Universe. And now the question is whether we live in a cyclic universe, where the grid of rings expands, expands with an acceleration and then contracts. Steinhardt and Turok worked with some sort of phantom field in their cyclic model of the universe from the year 2001, \cite{Steinh, Rosenzweig}. And so, could there be some connection between this phantom field and the scalar field, which is firmly built in the classical limit of RP, \cite{JNa6}? 

But that’s not the end of the list of interesting issues. The black hole information paradox still remains an important problem in cosmology, \cite{Preskill}. Basic suggestions of a solution to it point to a finding that the information simply disappears or is destroyed. We don't consider this explanation very credible. The second possibility is that there exists some remnant after the evaporation, where all the information in the black hole is stored. That could be one of the possible resolutions. We are above all excited by the idea that the superluminal signalling could be at the core of the argument about how information escapes from a black hole. That is exactly related to the fact that the gravitational rings mediate gravity superluminally, so in the final stage of this object's radiation, information would actually disappear along these objects that bridge the area inside the black hole and outside, \cite{JNa4}.

A fascinating area of research in quantum gravity phenomenology is the existence of dimensional reduction. As explained in the article \cite{Carlip} in most of the current approaches to quantum gravity, the dimension usually reduces to the dimension of 2. But why? We clearly see the reason in RP, because the ring is effectively a one-dimensional object on which we define the time. This means we obtain exactly a 2-dimensional theory in high-energy physics. But what is even more interesting is we could orient the fundamental ring (clockwise or counterclockwise). This issue is related to the existence of the arrow of time in the universe, where we assume that in an expanding universe time flows in one direction, but in a contracting universe the arrow of time reverses. But exploring this phenomenon will require further thermodynamic considerations, \cite{JNa4}.

RP brings us even more results by investigating the determinism of modern physical theories, \cite{Hooft}. The ongoing intensive debate in the quantum field theory community is what the role of probabilities is in this theory. RP brings a whole new light to it, because first the gravitational ring is created, on which is the elementary particle later travelling. Simply, the gravitational ring creates a trajectory on which particles later move. So, the Nature could be in our view extremely non-local, \cite{JNa4}. 

And what is really interesting and what inspired Penrose to the claim that quantum mechanics is incomplete is the problem of the collapse of the wave function, \cite{Singh}. On what physical scale does it occur, and what’s hidden behind it? RP offers actually a first hint of the solution: it will be the quantum gravity that gives us the answer because the gravitational ring decays to the quanta of graviton-phonons in the Planck time and "creates a trajectory for the movement of all particles and fields in the Universe", which means that the Nature behaves at the highest level of physics classically, with some built-in quantum behavior. Therefore, the probability description of the Nature at the quantum level won't be included in the fundamental description of laws. This is shocking news, \cite{JNa4}.

All of this is related to a philosophical trend in modern mathematics, which is called the finitism, \cite{Yamada}, when we reject all infinitely big quantities in modern physics. It follows from that that the Universe should not be infinitely big. It has a finite --- however extremely big --- diameter. We see it in the RP. Because the gravitational ring is created in the Planck time on arbitrary distances, and if the Universe were infinitely big, we would be back to the Newtonian limit that the gravitational interaction is mediated by infinitely big velocity. But we know that this could not be like that in the fundamental description of Nature. We assert that there are two velocities, the velocity of light c, which forms the ultimate speed of movement of all particles and fields, and the second velocity of $c_g>>10^{70}m/s$, which mediates gravity through a "massless" object - the gravitational ring, \cite{JNa1, JNa2, JNa4, JNa5, JNa6}.  So, we suggest the following generalization of the Lorentz transformations (in 2 dimensions, because the world is actually in high- energy sector according to RP 2-dimensional): 
\begin{gather}
t' = \frac{t - \frac{x}{v}\frac{v^2}{c^2}\epsilon - \frac{x}{v}\frac{v^2}{c_g^2}}{\sqrt{1 - \frac{v^2}{c^2}\epsilon - \frac{v^2}{c_g^2}} },\nonumber\\
x' = \frac{x - t v}{\sqrt{1 - \frac{v^2}{c^2}\epsilon - \frac{v^2}{c_g^2} }},
\end{gather}     
where ${\epsilon} = \epsilon(v)$ denotes some step function defined by the prescription
\begin{gather}
\epsilon(v) = \left\{
    \begin{array}{ll}
        1 & \mbox{for } v\leq c,\\
        0 & \mbox{for } v>c.
    \end{array}
\right.
 \end{gather}
The expression $\frac{v^2}{c_g^2}$ is very small for velocities $v < c < c_g$ and we obtain the standard Lorentz transformations in this case. But the given transformations have again the Lorentzian form for $c< v < c_g$. Only gravitational rings could be created by a velocity higher than c, and the ultimate goal is to glue the quantum field theory with the limiting velocity $c_g$ to the standard quantum field theory. 

Now, if we were to measure this velocity $c_g$, we would actually obtain the real diameter of the whole Universe, \cite{JNa4}. But this will not be so easy, because this velocity is beyond the sensitivity of all existing instruments and even all the instruments we will have available in the near future. But it does not mean that we could not test RP at the level of consistency. And first of all, we need to start examining what extension of the Poincar\'e group we have available today, \cite{Saviddy}. The next step will also be the generalization of the Haag-Lopuszanski-Sohnius theorem in mathematical physics, with which the whole program of supersymmetry actually started, \cite{JNa1}.  
 
\section*{Conclusion} 

We live in a fascinating time, where we are constantly refining the measurements of our models in cosmology. One of the key questions we have to ask is what the large-scale curvature of space in the universe is. All measurements so far have suggested that the Universe is spatially flat. But these observations don't rule out the possibility that the universe could have a geometry similar to that of a sphere or a saddle because of the error of the measurement. And the new results from the Planck satellite measurements suggest that it might actually be a geometry with a positive curvature parameter $k=1$.

But as sometimes happens, theory is ahead of experiment in its conclusions. It follows from the articles with fairly short calculations, \cite{Easson, EassonD}, that the curvature of the Universe is positive. Further, a recent article, \cite{JNa5}, came out, according to which it is possible to construct a very nonlocal theory of quantum gravity--- the ring paradigm. According to its basic postulate, the gravity is mediated by so-called gravitational rings, which are new phenomenological objects that decay into gravitons---which are phonons on the grid of matter in the Universe. This theory has many surprising implications for other areas of physics, but what’s important for our work is that it operates at two speeds: the speed of light c, which is the maximum speed of propagation for all Standard Model particles and fields---and for all particles, like supersymmetric ones, that we might potentially discover in the future---and the speed $c_g$, at which this gravitational ring is mediated. The velocity $c_g$ is the true maximal velocity of how the information is propagated in our Universe. But from that, after a bit of thinking, it clearly follows that the universe has a finite, though huge, diameter, because the speed at which gravity spreads must be finite. 

Already Hartle and Hawking pointed out this fact about curvature of the Universe in their Hartle-Hawking no-boundary proposal, \cite{Hawking}, which suggests our Universe has no initial spatial or temporal boundaries, meaning that time becomes spatial and the beginning of the cosmos is smooth rather than an infinite-density singularity. Their original theory doesn't work in an open or spatially flat universe! From the very first applications of RP to cosmology, it turns out that the preferred model is the model of cyclic universes, where the universe goes through a trillion-year cycle, expanding, accelerating in its expansion, and then swinging into the Big Crunch, \cite{JNa6}. Therefore, someone could argue that the results of the no-boundary proposal aren't so important to the applications of RP, because we work with different, the Wheeler-de Witt approach to quantum gravity, in it. But we can think it over that the cyclic models of the universe must have some initial cycle, otherwise we would be at odds with the results in thermodynamics, \cite{Steinh}. And therefore, in this initial cycle, the results of the original formulation of the no-boundary proposal, would be again relevant, which we all leave as a challenge for our future work. From that point of view, our thoughts about the possible curvature of the Universe take on a completely new shape.  

\section*{Acknowledgement}
We would like to thank Damien Easson for his comments. The research of M. K. was carried out in the Southern Federal University with financial support from the Ministry of Science and Higher Education of the Russian Federation (State contract FENW-2026-0028).

\end{document}